# MATRICES WHOSE INVERSIONS ARE TRIDIAGONAL, BAND OR BLOCK-TRIDIAGONAL AND THEIR RELATIONSHIP WITH THE COVARIANCE MATRICES OF A RANDOM MARKOV PROCESSES (FIELDS)

Ulan N. Brimkulov, Kyrgyz-Turkish Manas University, Bishkek, Kyrgyzstan

***Abstract*— The article discusses the matrices of the $A_n^1, A_n^m$, $A_N^m$ forms, whose inversions are: tridiagonal matrix $A_n^{-1}$ (*n* - dimension of the matrix), banded matrix $A_n^{-m}$ (*m* - the half-width band of the matrix) or block-tridiagonal matrix $A_N^{-m}$ (*N*=*n* x *m* – full dimension of the block matrix; *m* - the dimension of the blocks) and their relationships with the covariance matrices of measurements with ordinary (simple) Markov Random Processes (MRP), multiconnected MRP and vector MRP respectively. Such covariance matrices are frequently occurring in the problems of optimal filtering, extrapolation and interpolation of MRP and Markov Random Fields (MRF). It is shown, that the structure of the matrix $A_n^1, A_n^m$, $A_N^m$, has the same form, but the matrix elements in the first case are scalar quantities; in the second case matrix elements representing a product of vectors of dimension *m*; and in the third case, the off-diagonal elements are the product of matrices and vectors of dimension *m*. The properties of such matrices were investigated and a simple formulas of their inversion was founded. Also computational efficiency in the storage and inverse of such matrices have been considered. To illustrate the acquired results an example of the covariance matrix inversions of two-dimensional MRP is given.**

***Index Terms*— Best Linear Unbiased Estimates (BLUE), Markov process in the wide sense, simple (ordinary connected) Markov process, multiply connected (m-connected) Markov process, vector (m-dimensional) Markov process, Random field filtering and parametric identification, Tridiagonal Matrices, Banded Matrices and Block-Tridiagonal matrices.**

## I. INTRODUCTION

Research problems of random fields and processes may be founded in many applications, for instance, in the study of spatial and temporal variability of oceanographic fields (flow velocity fields, temperature fields and sea surface height, density and salinity fields, etc.), in the problems of statistical radio engineering and image fields reconstruction, and in many other engineering tasks.

The computing algorithms using the least-squares method (LSM), weighted and generalized LSM (WLSM, GLSM), Kalman-Bucy filter are usually used in estimation, filtering and interpolation of random fields based on the field realization measurements results. At the same time, if the investigated physical fields can be approximated by simple (ordinary connected), m-connected or vector Markov processes, we have used the computing schemes where we can meet tridiagonal, band and block-banded matrices. Therefore, in recent years, much attention is paid to the study of algorithms, computational efficiency, accounting the structure of the matrices included in the estimation algorithms, see e.g., [1]-[7].

In work [1] devoted to data assimilation in the study of large, multi-dimensional, time-dependent fields by accounting the structure of measurements matrix resulted in possibility to build four efficient Kalman-Bucy filter's algorithm, which reduces computing costs up to 2 orders, in comparison to known algorithms. This possibility appears in case when the measurement errors are approximated by Markov random field (MRF). In this case, the inverse covariance matrix of measurement errors field has a band structure that allows us to construct efficient in computational aspect algorithms. Accounting the sparse measurements, typical of tasks considered in the article (e.g., results of satellite scan) make it possible to obtain algorithms that are more efficient.

In [2] the matrices whose inversions are banded are considered. In this case, tridiagonal matrix represented as a Hadamard product of three matrices. This leads to very interesting result when Gauss-Markov's random process represented as the product of three independent processes: forward and backward processes with independent increments and a variance-stationary process. Here we can see the connection between matrices, entering the decomposition of the three-diagonal matrix and the processes entering the factorization of Gauss-Markov's random process. In this sense, the positive defined symmetric matrices with banded inverses can be viewed as a form of representation of Gauss-Markov's random processes. The paper also considered the problem of the approximation of the covariance matrix of non-stationary Gaussian process of general form by covariance matrix whose inversions are band matrix. The information loss of such approximation was estimated. This work also shows that for such matrix inversion it is necessary to know only the direct matrix elements, lying inside the band with the width L.

In [3] inversion algorithms of L-block banded matrices were obtained, their inversions are also L-block banded matrices. Received algorithms were applied to signal processing problems in the case of Kalman-Bucy Filtering (KBF) usage. These covariance matrices are approximated by block-band matrix. It gives an opportunity of reducing the computational complexity of the algorithms for 2 orders and makes the KFB algorithm feasible to solve problems of large dimension.

Summarizing the content of these papers we can note, that in all these works studied matrices, inversion of which leads to tridiagonal, banded or block-tridiagonal matrices. It is shown that if the matrices are symmetric and positive defined, then they can be covariance matrices of measurements of Gauss-Markov's random processes. The application of obtained results to signal processing tasks in the analysis of space-time random oceanographic fields were considered. The structure of these matrices allows obtaining efficient computational algorithms which let to solve large-scale tasks. The results are coupling to specific algorithms and programs and their effectiveness tested on examples of real experiments results processing.

Regardless of the results obtained in [1]-[7], the author in 1988-1992 obtained results that overlap with the results [1]-[7]. Unfortunately, these results are presented only in the form of manuscripts [18], [19] or published in the form of short abstracts of conferences in Russian (see., e.g., [21],[22]). There is only one article that was translated into English [20].

We can note, that the results related to the tridiagonal matrices inversion, the study of Markov processes were also considered in Russian mathematical literature, in literature on statistical radio engineering dedicated to various aspects of the random processes study [8]-[12]. But there are a few works in Russian literature that consider the use of banded and block-banded matrices in data processing algorithms. Also, the connection between Markov random processes (fields) class and covariance matrices of measurement of such fields ware poorly investigated. There is a few works on the evaluation of the effectiveness of computational processing circuits of Markov fields' research results. In [10] the relationship of Markov processes and stochastic differential equations in partial derivatives were shown and numerous examples of Markov processes were given.

## II. The Task of Optimum Estimating of Random Field's Mathematical Expectation

In many problems the analysis of random processes and fields based on experimental data (measurements), we are dealing with matrices, the dimensions of which increase with the number of measurements. That is, with increasing of the number of measurements the dimension of the matrix grows rapidly (in many cases proportional to the square of the measurements).

Consider one of the most common and well-known problems of the analysis of random processes (fields) - the problem of finding the best linear unbiased estimates (BLUE) of unknown parameters of the mathematical expectation model of the random field.

Let a random field $Z(t)$ describe by the model:

$$Z(t) = \eta(t) + \xi(t), \quad t \in T, \tag{1}$$

where $\eta(t) = \eta(t, \mathbf{B}) = \mathbf{f}^T(t)\mathbf{B}$ is mathematical expectation (deterministic component of the field), described by a linear-parameterized model with the vector of known linearly-independent functions $\mathbf{f}(t) = \left(f_1(t), \ldots, f_p(t)\right)^T$ and the vector of unknown parameters $\mathbf{B} = \left(B_1, \ldots, B_p\right)^T$; $\xi(t)$ - noise field (interference, measurement noise) with the known covariance function $k(s,t)$; $T$ - the interval, in which the model (1) is true.

Let us assume that there was set a problem: based on discrete measurement $Z(t)$ in the points $T_n = \{t_1 < t_n < \ldots < t_n \mid t_i \in T\}$ to find the Best Linear Unbiased Estimates (BLUE) $\hat{\mathbf{B}}_n$ parameters $\mathbf{B}$. The solving is well known and defined by the formula (see. e.g., [13,14]):

$$\hat{\mathbf{B}}_n = D_n F_n K_n^{-1} \mathbf{Z}_n \tag{2}$$

where

$$D_n = \left[F_n K_n^{-1} F_n^T\right]^{-1} \tag{3}$$

- covariance matrix of BLUE $\hat{\mathbf{B}}_n$; $F_n = [\mathbf{f}(t_1), \ldots, \mathbf{f}(t_n)]$ - matrix of vector $\mathbf{f}(t)$ values at the measurement points $T_n$; $\mathbf{Z}_n = \{Z(t_i) \mid t_i \in T_n\}$ - measurement vector; $K_n^{-1}$ - matrix is invers to the covariance matrix $K_n$ for $\mathbf{Z}_n$: $K_n = \{k(t_i, t_j)\} = \{k_{ij}\}$ $(i, j = \overline{1, n})$ $(t_i \in T_n)$.

In spite of optimality of estimations (2), the use under large number of measurements becomes difficult or impossible. This is because of the fact that in the expressions (2) and (3) enters the matrix $K_n^{-1}$, which number of elements is increasing in proportion to the measurements square. The $K_n^{-1}$ matrix computing and storage requires large computational costs (required memory in proportion to $n^2$, and the number of operations for the $K_n$ matrix inversion in proportion to $n^3$ [17]).

The same matrix $K_n^{-1}$ is a part of expression to calculate the optimal estimates for more general tasks of processing of random processes based on the measurement results. For example, in the formulas of filtering, interpolation and extrapolation of a random field based on the measurement results is also included matrix $K_n^{-1}$. The weighting matrix of size $(n \times n)$, is also included in the evaluation formula, when used generalized least-squares method (GLSM). We do not present here the well-known results, which can be found in numerous literatures, see e.g. [13], [14].

So, here is a rather actual problem of finding of a class of random processes (fields) for which the covariance matrix of the measurement $K_n$ is such, that its inverse matrix $K_n^{-1}$ is sparse and in particular has the tridiagonal, band or block-tridiagonal structure. In this case, the calculation of optimal estimates, including BLUE for many problems of random processes and fields analysis is greatly simplified from a computational point of view by taking into account the structure of the covariance matrix measurement.

## III. Matrices Whose Inversions are Tridiagonal, Band and Block-Tridiagonal

Below we shall consider three classes of square matrices, whose inversion the tridiagonal, band and block-tridiagonal matrices. This may be noticed, that all of these three classes of matrices has the similar structures but has some differences:

- in the first case, the matrix elements are formed from a scalar quantity;
- in the second case, the elements of the matrix is the result of multiplication of matrix and vector quantities;
- in the third case, the elements of the matrix formed from the blocks which are the product of square matrices of smaller dimension.

*A. A class of matrices, whose inversion leads to tridiagonal matrices*

Let matrix $A = A_n^1$ (size *n*x*n*) has the following form:

$$A_n^1 = \begin{bmatrix} a_{11} & \Lambda_{11}a_{11} & \Lambda_{12}a_{11} & \cdot & \cdot & \cdot & \Lambda_{1,n-1}a_{11} \\ \Gamma_{11}a_{11} & a_{22} & \Lambda_{22}a_{22} & \cdot & \cdot & \cdot & \Lambda_{2,n-1}a_{22} \\ \Gamma_{21}a_{11} & \Gamma_{22}a_{22} & a_{33} & & \cdot & & \cdot \\ \cdot & \cdot & & \cdot & & \cdot & \cdot \\ \cdot & \cdot & \cdot & & \cdot & & \cdot \\ \cdot & \cdot & & \cdot & & a_{n-1,n-1} & \Lambda_{n-1,n-1}a_{n-1,n-1} \\ \Gamma_{n-1,1}a_{11} & \Gamma_{n-1,2}a_{22} & \cdot & \cdot & \cdot & \Gamma_{n-1,n-1}a_{n-1,n-1} & a_{nn} \end{bmatrix} \tag{4}$$

where $\Gamma_{ij} = \prod_{l=j}^{i} \gamma_l \, (i \ge j)$ and $\Lambda_{ij} = \prod_{l=i}^{j} \lambda_l \, (j \ge i)$ $(i, j = 1, 2, \ldots, n-1)$, $a_{ii}$ $(i = 1,2,\ldots,n)$, $\gamma_i$, $\lambda_i$ $(i = 1,2,\ldots,n-1)$ – arbitrary real numbers.

Thus, the off-diagonal elements $A_n^1$ are determined by the expression

$$a_{ij} = \begin{cases} \Gamma_{j,i-1} a_{jj} = a_{jj} \prod_{l=j}^{i-1} \gamma_l & if \quad i > j \\ \Lambda_{i,j-1} a_{ii} = a_{ii} \prod_{l=i}^{j-1} \lambda_l & if \quad j > i \end{cases} \qquad (i, j = \overline{1, n-1}) \tag{5}$$

Here we can formulate the following

*Theorem 1:* Let the matrix $A_n$ has the form (4) and $\det A_n \neq 0$, then

1. Matrix $A_n^{-1}$, inverse to (4) will have tridiagonal form:

$$A_n^{-1} = \begin{bmatrix} \frac{\mu_1}{\alpha_1\alpha_2} & -\frac{\lambda_1}{\alpha_2} & 0 & \cdot & \cdot & \cdot & 0 \\ -\frac{\gamma_1}{\alpha_2} & \frac{\mu_2}{\alpha_2\alpha_3} & -\frac{\lambda_2}{\alpha_3} & 0 & \cdot & \cdot & \cdot \\ 0 & -\frac{\gamma_2}{\alpha_3} & \cdot & & \cdot & & \cdot \\ \cdot & 0 & \cdot & \cdot & \cdot & \cdot & \cdot \\ \cdot & \cdot & & \cdot & \cdot & \cdot & 0 \\ \cdot & \cdot & \cdot & & \cdot & \frac{\mu_{n-1}}{\alpha_{n-1}\alpha_n} & -\frac{\lambda_{n-1}}{\alpha_n} \\ 0 & \cdot & \cdot & \cdot & 0 & -\frac{\gamma_{n-1}}{\alpha_n} & \frac{1}{\alpha_n} \end{bmatrix}, \tag{6}$$

where $\alpha_i = a_{ii} - \gamma_{i-1}\lambda_{i-1}a_{i-1,i-1}$ $(i=\overline{2,n})$, $\alpha_1 = a_{11}$, $\mu_i = a_{i+1,i+1} - \gamma_{i-1}\gamma_i\lambda_{i-1}\lambda_i a_{i-1,i-1}$ $(i=\overline{2,n-1})$, $\mu_1 = a_{22}$.

2. The determinant of any corner of the submatrix $A_i^1$, including the determinant of the complete matrix $A_n^1$, can be calculated by the expression

$$\det A_i^1 = \prod_{l=1}^{i} \alpha_l \,,\ (i=\overline{1,n}).$$

This theorem proofing uses the method of induction and recursive procedure of matrix inversion by method of step-by-step bordering can be found in Appendix.

*Note 1.* The matrix (4) can be also written as

$$A_n^1 = \begin{bmatrix} a_{11} & \lambda_1 a_{11} & \lambda_2 a_{12} & \cdot & \cdot & \cdot & \lambda_{n-1}a_{1,n-1} \\ \gamma_1 a_{11} & a_{22} & \lambda_2 a_{22} & \cdot & \cdot & \cdot & \lambda_{n-1}a_{2,n-1} \\ \gamma_2 a_{21} & \gamma_2 a_{22} & a_{33} & & \cdot & & \cdot \\ \cdot & \cdot & & \cdot & & \cdot & \cdot \\ \cdot & \cdot & \cdot & & \cdot & & \cdot \\ \cdot & \cdot & & \cdot & & a_{n-1,n-1} & \lambda_{n-1}a_{n-1,n-1} \\ \gamma_{n-1}a_{n-1,1} & \gamma_{n-1}a_{n-1,2} & \cdot & \cdot & \cdot & \gamma_{n-1}a_{n-1,n-1} & a_{nn} \end{bmatrix} \qquad (7)$$

where $a_{ii}(i=\overline{1,n})$, $\gamma_i$, $\lambda_i$ ($i$ = 1, 2,…, $n$-1) - arbitrary real numbers, $a_{ij}(i=\overline{1,n-1}\,||\, i \neq j)$ - determined in (5). Expressing $a_{31}(a_{13})$ through $a_{21}(a_{12})$, $a_{41}(a_{14})$ through $a_{31}(a_{13})$, etc., the matrix (7) can be presented to the form (4).

*Note 2.* The Matrix (4) is completely determined by $3n-2$ elements, which are included into the three central diagonals $A_n^1$. Other elements of $A_n^1$ cancel each other while inversion.

*Note 3*. The Matrix (4) (or (7)) is only one form of the square matrices representation, whose inverse are tridiagonal matrices. Many papers devoted to matrices research questions, whose inversions are tridiagonal matrices (see, for instance, [15] and other). The $A_n^1$ matrices recording in the form of (4) (or (7)) is convenient because the results are easily generalized to the matrices whose inversion are banded or block-tridiagonal matrices.

*B. The class of matrices whose inversion leads to banded matrices with the half-band's width m.*

The results of Theorem 1 can be generalized to the band matrices with $1 \le m \le n - 1$, where $n$ - dimension of the reversible matrix; $m$ - the half-width band of band inverse matrix.

To formulate the main theorem of this section, a number of new notations will be required. We partition up (divide) the square matrix $A_i$ ($i$ - order of matrix) on the sub-matrix as follows:

$$A_i = \begin{array}{c} \\ \\ i-m\left\{\begin{array}{c}1\\2\\ \vdots \\ i-m\end{array}\right. \\ m\left\{\begin{array}{c} i-m+1 \\ \vdots \\ i \end{array}\right. \end{array} \begin{array}{c} \overbrace{1 \;\; 2 \;\; \cdots \;\; i-m}^{i-m} \; \overbrace{i-m+1 \;\; \cdots \;\; i}^{m} \\ \begin{Vmatrix} A_{i-m} & A_{im} \\ A_{mi} & A_m[i] \end{Vmatrix} \end{array} \qquad (8)$$

where $A_{im}$ - the sub-matrix with the $i \times m$ size, representing right $m$ – columns of the matrix $A_i$; $A_{mi}$ - sub-matrix with the $mxi$ size, representing a lower m-rows of $A_i$ ; $A_{i-m}$ - $(i-m)\times(i-m)$ left upper diagonal submatrix $A_i$; $A_m[i]$ - ($m$ x $m$) lower right diagonal submatrix $A_i$.

*Note 4.* 1. Notation [$i$] shows, that index $i$ changes from $i$-$m$+1 to $i$. If $i < m$, then index $i$ changes from 1 to $i$.

2. For $i \le m$ sub-matrix $A_{i-m} = [0]$, sub-matrices $A_{im}$ , $A_{mi}$ and $A_m[i]$ coincide with each other and has sizes ($i$ x $i$).

Let's consider the trapezoidal real matrices $\Im_{n-1,m}$ and $\Re_{m,n-1}$:

$$\Im_{n-1,m} = \begin{Vmatrix} & & & & \gamma_{1m} \\ & & & \gamma_{2,m-1} & \gamma_{2m} \\ & & \iddots & \vdots & \vdots \\ & \gamma_{m-1,2} & \cdots & \gamma_{m-1,m-1} & \gamma_{m-1,m} \\ \gamma_{m1} & \gamma_{m2} & & \gamma_{m,m-1} & \gamma_{mm} \\ \vdots & \vdots & & \vdots & \vdots \\ \gamma_{n-1,1} & \gamma_{n-1,2} & \cdots & \gamma_{n-1,m-1} & \gamma_{n-1,m} \\ 1 & 2 & \ldots & m\text{-}1 & m \end{Vmatrix} = \begin{Vmatrix} \boldsymbol{\gamma}^T_{1m} \\ \boldsymbol{\gamma}^T_{2m} \\ \vdots \\ \boldsymbol{\gamma}^T_{m-1,m} \\ \boldsymbol{\gamma}^T_{mm} \\ \vdots \\ \boldsymbol{\gamma}^T_{n-1,m} \end{Vmatrix} \begin{matrix} 1 \\ 2 \\ \vdots \\ m\text{-}1 \\ m \\ \vdots \\ n\text{-}1 \end{matrix} \qquad (9)$$

and

$$\Re_{m,n-1} = \begin{Vmatrix} & & & & \lambda_{1,m} & \cdots & \lambda_{1,n-1} \\ & & & \lambda_{2,m-1} & \lambda_{2,m} & & \lambda_{2,n-1} \\ & & & \vdots & \vdots & & \vdots \\ & \lambda_{m-1,2} & \cdots & \lambda_{m-1,m-1} & \lambda_{m-1,m} & \cdots & \lambda_{m-1,n-1} \\ \lambda_{m1} & \lambda_{m2} & \cdots & \lambda_{m,m-1} & \lambda_{mm} & \cdots & \lambda_{m,n-1} \end{Vmatrix} \begin{matrix} 1 \\ 2 \\ \vdots \\ m\text{-}1 \\ m \end{matrix} \qquad (10)$$

$$= \begin{Vmatrix} \boldsymbol{\lambda}_{m1} & \boldsymbol{\lambda}_{m2} & \cdots & \boldsymbol{\lambda}_{m,m-1} & \boldsymbol{\lambda}_{mm} & \cdots & \boldsymbol{\lambda}_{m,n-1} \end{Vmatrix}$$
$$\begin{matrix} 1 & 2 & \cdots & m\text{-}1 & m & \cdots & n\text{-}1 \end{matrix}$$

*Note 5.* 1. Matrices $\Im_{n-1,m}$ and $\Re_{m,n-1}$ has *m*(*n*-(*m*+1)/2) elements.

2. Vector-rows $\boldsymbol{\gamma}_{im}^T$ and vector-columns $\boldsymbol{\lambda}_{mi}$ for $m \le i \le n-1$ has fixed length, which equal to *m*. For $i < m$ vectors $\boldsymbol{\gamma}_{im}^T$ and $\boldsymbol{\lambda}_{mi}$ has variable length, which equal to $i$ $(1 \le i < m)$.

Let the square matrix $A_n = A_n^m$ of $n$ order forms as follows:

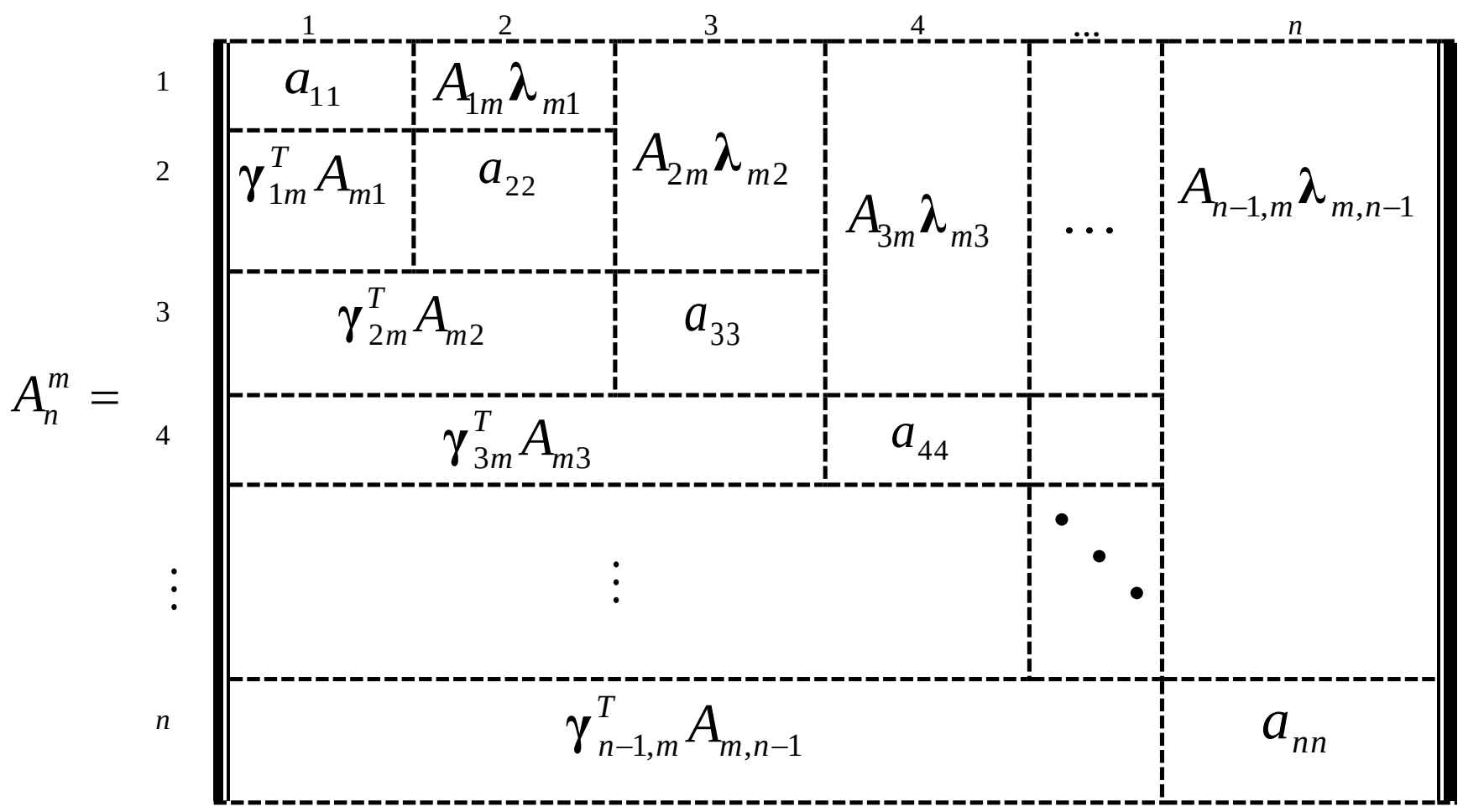

(11)

where $a_{ii} (i = \overline{1,n})$ - arbitrary real numbers; $\boldsymbol{\gamma}_{im}^T$, $\boldsymbol{\lambda}_{mi}$ $(i = \overline{1, n-1})$ - real vectors, with the length $i$ for $(i \le m)$ and $m$ for $m < i \le n-1$; (sub)matrix $A_{mi}$ and $A_{mi}$ determined in (8).

In view of these notations, we can formulate

*Theorem 2:* Suppose that the matrix $A_n = A_n^m$ has the form (11), then, if $\det A_n^m \ne 0$, the following statements are true:

1. The matrix $A_n^{-m} = \left[A_n^m\right]^{-1} = \left[c_{ij}\right]_{i,j=1}^n$, has the band form with the half-width $m$ $(1 \le m \le n-1)$, i.e. elements of $A_n^{-m}$ satisfy the condition: $c_{ij} = 0$ for $|i - j| > m$.

2. The non-zero elements $A_n^{-m}$, lying inside the band which half-width equals *m* can be found as follows:

a) calculating of auxiliary quantities $\{\alpha_i, (i = \overline{1,n})\}$:

$$\alpha_i = a_{ii} - \boldsymbol{\gamma}_{i-1,m}^T A_m[i-1]\boldsymbol{\lambda}_{m,i-1}, (i = \overline{1,n}), \tag{12}$$

where $\boldsymbol{\gamma}_{0m}^T = \boldsymbol{\lambda}_{m0} = A_m[0] = 0$.

b) calculating of diagonal elements $\{c_{ii}, i = \overline{1,n}\}$:

$$c_{ii} = \frac{1}{\alpha_i} + \sum_{k=0}^{w} \frac{\lambda_{m-k,i+k}\gamma_{i+k,m-k}}{\alpha_{i+k+1}}, \tag{13}$$

where $w = m - 1$ if $i \le n - m$ and $w = n - i - 1$ if $i > n$-$m$.

c) calculating of off-diagonal elements of upper $\{c_{i,i+k}, k = \overline{1,m}\}$ and lower $\{c_{i+k,i}, k = \overline{1,m}\}$ half-bands $(i = \overline{1, n-1})$:

$$c_{i+k,i} = -\frac{\gamma_{i+k-1,m-k+1}}{\alpha_{i+k}} + \sum_{j=k}^{w} \frac{\lambda_{m+k-j,i+j}\gamma_{i+j,m-j}}{\alpha_{i+j+1}}, \tag{14}$$

$$c_{i,i+k} = -\frac{\lambda_{m-k+1,i+k-1}}{\alpha_{i+k}} + \sum_{j=k}^{w} \frac{\lambda_{m-j,i+j}\gamma_{i+j,m+k-j}}{\alpha_{i+j+1}} \tag{15}$$

where $w$ defined similarly as $w$ in (13).

In (13)-(15), if the calculated value of upper limit becomes smaller than the lower limit, the summarizing should not be executed, i.e. the second term on the right part of formulas (13)- (15) for $w < k$ is identically equal to 0.

3. The determinant of any corner submatrix $A_i^m \quad (i = \overline{1,n})$, including the determinant of the complete (full) matrix $A_n^m$, can be calculated using the expression:

$$\det A_i^m = \prod_{l=1}^{i} \alpha_l \ , \ (i = \overline{1,n}).$$

*Note 6.* The matrix $A_n^m$ can be written in a form similar to (7) with the replacement of the off-diagonal scalar elements $a_{ij}$ $(i,j = \overline{1,n-1 \mid i \neq j}))$ and $\gamma_i, \lambda_i (i = \overline{1,n-1})$ to vectors $\mathbf{a}_{[i],j} (i = \overline{1,n-1}; j = \overline{i-1,n-1})$, $\mathbf{a}_{i,[j]}^T (i = \overline{j-1,n-1; j = n-1})$ and $\boldsymbol{\gamma}_i^T, \boldsymbol{\lambda}_i (i = \overline{1,n-1})$, appropriately, having a length $i$ for $i \le m$ and length $m$ for $m < i \le n-1$.

Taking into account the note 6 of theorem 2 proofing fully repeats the theorem 1 proofing with the replacement of scalar values

$\gamma_i, \lambda_i \ (i = \overline{1,n-1})$ and $a_{ij}, a_{ji} (i,j = \overline{1,n-1} \mid i \neq j)$ with vectors

$\boldsymbol{\gamma}_i^T, \boldsymbol{\lambda}_i (i = \overline{1,n-1})$ and $\mathbf{a}_{[i]j}, \mathbf{a}_{i[j]}^T (i,j = \overline{1,n-1} \mid i \neq j)$.

*Note 7.* From formulas (12) - (15) show, for the matrix inversion of the form $A_n^m$ it is enough to know it's elements, lying inside the band with a width of 2*m*+1. Other elements of $A_n^m$ cancel each other during inversion. In other words, the matrix $A_n^m$ is completely determined by its elements lying inside the band width 2*m*+1.

*Note 8.* The matrix $A_n^m$ depends on $w^* = (2m+1)n - m(m+1)$ of arbitrary selected values $\{a_{ii}, i = \overline{1,n}\}$, $\{\gamma_i^*, \lambda_i^*, i = \overline{1,w}\}$, or, in other words, has $w^*$ independent elements. Other elements of $A_n^m$ linear connected with them.

If $m = n-1$, the number of independent values which depend on the elements of the matrix $A_n^m$ becomes equal to $n^2$, i.e. we come to a matrix of general form $(A_n^m = A_n^{n-1} = A_n)$. In this case, the inverse matrix will be completely filled, i.e. it will have $n^2$ elements.

If *m* = 1, then value $w^* = 3n - 2$. For $A_n^m = A_n^1$ and inverse matrix $A_n^{-m} = A_n^{-1}$ will be tridiagonal. For $m = 0$ matrix $A_n^m$ = $A_n^0$ and inverse for it matrix $\left(A_n^0\right)^{-1}$ will be diagonal.

*Note 9.* Matrices of $A_n^m$ form can be stored in memory in a compact form. It is sufficient to introduce in memory the vectors $\mathbf{a}_{ii} = \{a_{ii}, (i = \overline{1,n})\}$, $\boldsymbol{\gamma}_w^*$ and $\boldsymbol{\lambda}_w^*$, i.e. of matrices $\Im_{n-1,m}, \Re_{m,n-1}$, holding $w^* \le n^2$ memory cells. With the help of vectors $\mathbf{a}_{ii}$, $\boldsymbol{\gamma}_w^*$ and $\boldsymbol{\lambda}_w^*$ if necessary, it can be easily calculated every element of the matrix $A_n^m$. Thus, if $m << n$, the gain in the amount of required memory can be considerable (significant).

*C. Matrices class, whose inversion leads to a block tridiagonal matrix.*

Let the matrix $A_N$ with the size $(N \times N)$ have a form:

$A_N = \mathbf{A}_N^m = \left[A_{ij}\right]_{i,j=1}^n$, where $A_{ij}$ - square sub-matrixes (blocks) with the size $(m \times m)$; $(N = n \times m)$.

Let non-diagonal sub-matrices $A_n^m$ determined by following expressions (compare with (5)):

$$A_{ij} = \begin{cases} \boldsymbol{\Gamma}_{j,i-1} A_{jj} = \left[\prod_{l=j}^{i-1} \Gamma_l\right] A_{jj} & if \quad i \ge j \ , \\ A_{ii} \boldsymbol{\Lambda}_{i,j-1} = A_{ii} \prod_{l=i}^{j-1} \Lambda_l & if \quad j \ge i \ , \end{cases} \quad (i,j = \overline{1,n-1}) \tag{16}$$

where $\Lambda_i \quad (i = \overline{1,n-1})$ and $\Gamma_i \quad (i = \overline{1,n-1})$ - matrix of real elements of the size $(m \times m)$; square (sub)matrices $\boldsymbol{\Lambda}_{ij}$ and $\boldsymbol{\Gamma}_{ij} (i,j = \overline{1,n-1})$ of *n* order can also be written as follows:

$$\begin{cases} \mathbf{\Gamma}_{ij} = \Gamma_j \Gamma_{j+1} \dots \Gamma_{i-1}\Gamma_i \quad if \quad i \ge j \\ \mathbf{\Lambda}_{ij} = \Lambda_i \Lambda_{i+1} \dots \Lambda_{j-1}\Lambda_j \quad if \quad j \ge i \end{cases} \quad (i, j = \overline{1, n-1}). \tag{17}$$

Let the matrix $A = A_N^m$ has a form:

$$\mathbf{A}_N^m = \begin{Vmatrix} A_{11} & A_{11}\mathbf{\Lambda}_{11} & A_{11}\mathbf{\Lambda}_{12} & \cdots & A_{11}\mathbf{\Lambda}_{1,n-1} \\ \mathbf{\Gamma}_{11}A_{11} & A_{22} & A_{22}\mathbf{\Lambda}_{22} & \cdots & A_{22}\mathbf{\Lambda}_{2,n-1} \\ \mathbf{\Gamma}_{21}A_{11} & \mathbf{\Gamma}_{22}A_{22} & A_{33} & & \vdots \\ \vdots & \vdots & & A_{n-1,n-1} & A_{n-1,n-1}\mathbf{\Lambda}_{n-1,n-1} \\ \mathbf{\Gamma}_{n-1,1}A_{11} & \mathbf{\Gamma}_{n-1,2}A_{22} & \cdots & \mathbf{\Gamma}_{n-1,n-1}A_{n-1,n-1} & A_{nn} \end{Vmatrix} \tag{18}$$

Elements (blocks) (18) depend only on $(3n-2)m^2$ scalar quantities constituting sub-matrices $A_{ii}(i = \overline{1, n})$, $\mathbf{\Lambda}_{ij}$ and $\mathbf{\Gamma}_{ij}(i, j = \overline{1, n-1})$.

Taking into account given notations we can formulate the following

*Theorem 3*: 1) Matrix $\mathbf{C}_N^m = \left(\mathbf{A}_N^m\right)^{-1} = \mathbf{A}_N^{-m}$, divided on $(m \times m)$ sub matrices $C_{ij}^m$, has banded-tridiagonal form with non-zero blocks:

$$C_{ii}^m = \Omega_{i+1}^{-1}\Sigma_i\Omega_i^{-1}, \quad (i = \overline{1, n-1}), \quad C_{nn}^m = \Omega_n^{-1}, \tag{19}$$

$$C_{i,i+1}^m = -\Lambda_i\Omega_{i+1}^{-1}, \quad C_{i+1,i}^m = -\Gamma_i\Omega_{i+1}^{-1} \quad , \tag{20}$$

where

$$\Omega_i = A_{ii} - \Gamma_{i-1}A_{i-1,i-1}\Lambda_{i-1}, \quad (i = \overline{2, n}), \quad \Omega_1 = A_{11}, \tag{21}$$

$$\Sigma_i = A_{i+1,i+1} - \Gamma_i\Gamma_{i-1}A_{i-1,i-1}\Lambda_{i-1}\Lambda_i, \quad (i = \overline{2, n-1}), \quad \Sigma_1 = A_{22}. \tag{22}$$

So, the general form of matrix $\mathbf{C}_N^m = \mathbf{A}_N^{-m}$ will have the following form:

$$\mathbf{C}_N^m = \begin{Vmatrix} \Omega_2^{-1}\Sigma_1\Omega_1^{-1} & -\Gamma_1\Omega_2^{-1} & & & & \\ -\Omega_2^{-1}\Lambda_1 & \Omega_3^{-1}\Sigma_2\Omega_2^{-1} & -\Gamma_2\Omega_3^{-1} & & \mathbf{0} & \\ & -\Omega_3^{-1}\Lambda_2 & \Omega_4^{-1}\Sigma_3\Omega_3^{-1} & \ddots & & \\ & & -\Omega_4^{-1}\Lambda_3 & & & \\ & \mathbf{0} & & & \Omega_n^{-1}\Sigma_{n-1}\Omega_{n-1}^{-1} & -\Gamma_{n-1}\Omega_n^{-1} \\ & & & & -\Omega_n^{-1}\Lambda_{n-1} & \Omega_n^{-1} \end{Vmatrix}$$

3) The determinant of any corner submatrix $\mathbf{A}_i^m(i = \overline{1, n})$, including the determinant of the complete matrix $\mathbf{A}_N^m$ can be calculated by the expression

$$\det \mathbf{A}_i^m = \prod_{l=1}^{i} \det \Omega_l .$$

The theorem 3 proofing repeats the theorem 1 proofing with the replacement of scalar quantities in (4) in the sub-matrices (18).

For theorem 3 all the notes which were given to theorems 1 and 2 are also true, bearing in mind that we are dealing with a block matrix.

## IV. Reducing of Operations Number and Required Memory for Inversion of A Symmetric Matrix of the Form $\mathbf{A}_N^m$ Relatively to General Matrix

In the formation of (sub) matrices $\mathbf{A}_N^{-m}$ in the symmetric case, only 2*n*-1 blocks of the matrix $\mathbf{A}_N^m$ are included being the elements of its main diagonal, and one of the adjacent side of the diagonals. Thus, in the computer memory sufficient to store $(2n\text{-}1)m^2$ quantities constituting mentioned submatrices $\mathbf{A}_N^m$, that at large $N$ will give a substantial savings in computer memory. In the case of the covariance matrix of general form in a computer memory it is necessary to store $N(N+1)/2$ values.

For calculation of non-zero submatrices $C_{ii}$ $(i=\overline{1,n})$ and $C_{i,i+1}$ $(i=\overline{1,n-1})$ the matrix $\mathbf{A}_N^{-m}$, it is necessary to calculate previously (*n*-1) submatrices $\Omega_i$ $(i=\overline{2,n})$ (submatrix $\Omega_1 = A_{11}$) and (*n*-2) submatrices $\Sigma_i$ $(i=\overline{2,n-1})$ (submatrix $\Sigma_1 = A_{22}$).

Calculations show that the required number of operations such as multiplication on submatrices calculation is equal to:

1) $\Omega_i \quad (i=\overline{1,n}) \quad \Rightarrow \quad (n-1)\left[m^3+m^2(m+1)/2\right]$;

2) $\Sigma_i \quad (i=\overline{2,n-1}) \quad \Rightarrow \quad (n-2)\left[m^3+m^2(m+1)/2\right]$;

3) $C_{ii} \quad (i=\overline{1,n}) \quad \Rightarrow \quad (n-1)\left[m^2(m+1)\right]$;

4) $C_{i,i+1} \quad (i=\overline{1,n-1}) \quad \Rightarrow \quad (n-1)m^2$.

Thus, the calculation of the matrix $\mathbf{A}_N^{-m}$ requires only $nm^2(4m+3)-m^2(11m+7)/2$ multiplications plus $\approx nm^2$ operations on $\Omega_i$ $(i=\overline{1,n})$ submatrices inversion. The last expression shows that the number of multiplications is proportional to *n* and $m^3$.

By calculating the exact number of additions and subtractions required for inversion of $\mathbf{A}_N^m$, obtain an expression $(5n-6{,}5)m^3-(2n-2{,}5)m^2+(n-1)m$, which is also proportional to *n* and $m^3$.

It is known [17], that the number of arithmetic operations required for general matrix inversion with the size $(n\times m)\times(n\times m)$ is proportional to $(n \text{ x } m)^3$. Thus, for large values of the ratio *n*/*m* (that usually takes place in the tasks under study), accounting the structure of the covariance matrix of the observed Markov process gives an opportunity to simplify the calculation of the required estimates.

The ratio of the number of non-zero elements of the symmetric matrix $\mathbf{A}_N^m$ to the number of the elements of the filled symmetric matrix with different values of *n* and *m* is shown in Table 1. From the Table 1 it is seen that the gain in the required memory amount practically independent of *m* and proportional to *n*.

Fig. 1 shows memory saving graphs and the number of arithmetic operations in dependence of *n* and *m* for matrix $\mathbf{A}_N^m$ inversion.

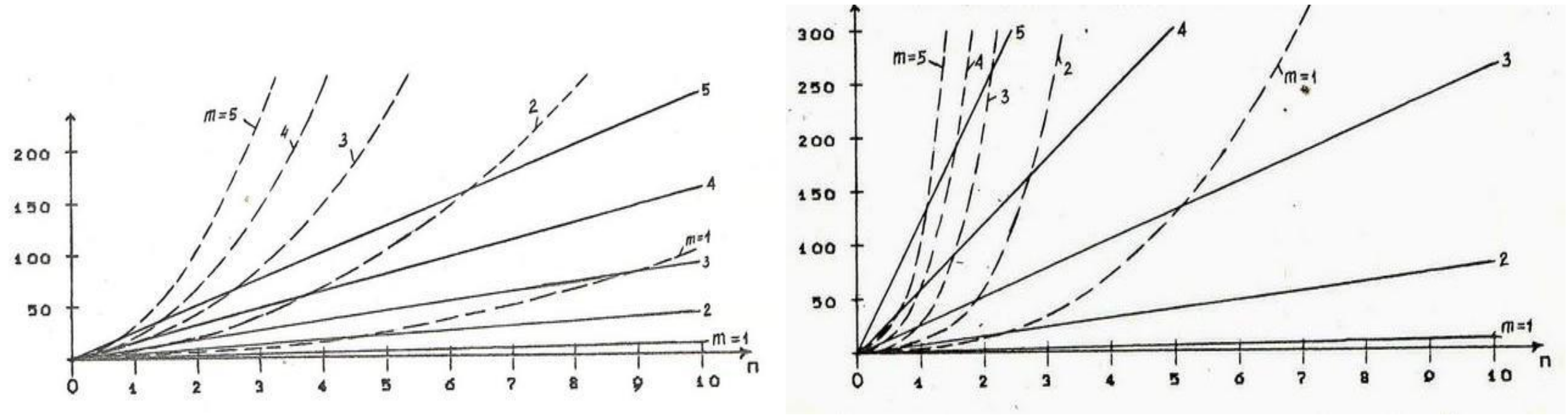

(a) required memory: $(nm)^2 \rightarrow nm^2$ (b) the number of arithmetical operations $(nm)^3 \rightarrow nm^3$

Fig. 1. The required amount of memory (a) and the number of arithmetic operations (b), which is necessary for inversion of the matrix of the form $\mathbf{A}_N^m$ (continuous lines) and general matrices (dashed lines).

TABLE I

THE RATIO OF THE NUMBER OF ELEMENTS OF THE FILLED MATRIX $(N \times N)$ TO THE NUMBER OF NON-ZERO ELEMENTS OF THE MATRIX $\mathbf{A}_N^m$ WITH AN ALLOWANCE OF THEIR SYMMETRY:

| m \ n | 5 | 10 | 50 | 100 | 500 | 1000 |
|---|---|---|---|---|---|---|
| 1 | 1.67 | 2.89 | 12.88 | 25.38 | 125.38 | 250.38 |
| 2 | 1.53 | 2.76 | 12.75 | 25.25 | 125.25 | 250.25 |
| 3 | 1.48 | 2.72 | 12.71 | 25.21 | 125.21 | 250.21 |
| 4 | 1.46 | 2.70 | 12.69 | 25.19 | 125.19 | 250.19 |
| 5 | 1.44 | 2.68 | 12.68 | 25.18 | 125.18 | 250.18 |

## V. THE RELATIONSHIPS OF MATRICES $A_n^1$, $A_n^m$, $\mathbf{A}_N^m$ WITH THE COVARIANCE MATRICES OF MEASUREMENTS OF ORDINARY, M - CONNECTED AND VECTOR MARKOV PROCESSES

### A. *The covariance matrix of Markov process measurements in a wide-sense*

Let the observed process $Z(t)$ is a Markov process in the wide-sense (further instead of "Markov process in the wide-sense" we will simply write "Markov process"). This means that the covariance function $k(s,t)$ of the process $Z(t)$ satisfies the condition [8]

$$k(s,t) = k(s,\tau)k(\tau,t)/k(\tau,\tau) \quad (s < \tau < t) \tag{23}$$

From the Doob Theorem [8] follows, that the condition (23) is not only necessary but also sufficient, i.e. positive definite function $k(s,t)$ is the covariance function of a Markov process only in case when it satisfies the condition (23).

Let the values of $\gamma_i$ ($i = 1,2,\ldots$) define as follows:

$$\gamma_i = k_{i,i+1}/k_{ii} \tag{24}$$

where $k_{ij} = k(t_i, t_j)$ – values of the covariance function of process $Z(t)$ at the points $t_i$ and $t_j$, $t_j > t_i$. Thus, $\gamma_i$ are the coefficients of the covariance of neighboring points reduced to a dispersion quantity in the points with a lower coordinate value. For stationary random processes $\gamma_i = \rho_{i,i+1}$, i.e. $\gamma_i$ - correlation coefficients between adjacent measurement process.

Taking into account (23) and (24) we can formulate the following

*Theorem 4:* 1) Covariance matrix $K_n$ of Markov process $Z_M(t)$ measurements at points $T_n = \{t_1 < t_2 < \ldots < t_n \mid t_i \in T\}$ is a special case of the matrix (4) for $\Lambda_{ji} = \Gamma_{ij} = \prod_{l=j}^{i} \gamma_l (i \geq j), (i, j = \overline{1, n-1})$, where $\gamma_i$ – previously defined in (24).

2) Elements of $K_n^{-1}$ defined by expression (6) taking into account that $\lambda_i = \gamma_i \;\; (i = \overline{1, n-1})$ and $a_{ii} = k_{ii} \; (i = \overline{1,n})$. At the same time

$$\alpha_i = k_{ii} - \gamma_{i-1}^2 k_{i-1,i-1} \quad (i = \overline{2,n}), \quad \alpha_1 = k_{11}, \tag{25}$$

$$\mu_i = k_{i+1,i+1} - \gamma_{i-1}^2 \gamma_i^2 k_{i-1,i-1} \quad (i = \overline{2, n-1}), \quad \mu_1 = k_{22}. \tag{26}$$

*Theorem 4' (inverse).* Any symmetric positive defined matrix of the form $A_n^1$ is the covariance matrix of the Markov process measurements.

The theorems 4 and 4' proofing are given in [20] (see in [20] Theorem 2 and note 3).

*Note 10.* It is follows from the note 2, that the matrix $K_n^1$ is completely determined by the elements of its two diagonals (the main diagonal and parallel to it, above or below). In other words, *the matrix $K_n^1$ depends only on the dispersion (variance) values in the measuring n points and the (n – 1) coefficients of the covariance between adjacent measurement points.*

The covariance matrix of the measurements of the Markov process is completely determined by small number ($2n$-1) of its elements. Therefore, for effective solving the Markov random processes of statistics problems is sufficient only a priori knowledge of the mentioned elements of the covariance matrix.

### B. *The covariance function and the covariance matrix of measurement of m-connected Markov process*

Let $Z(t)$ be an $m$-connected Markov process. This means that the covariance's between discrete measurements of the process $Z(t)$ satisfies to the condition

$$k(t_i, t_j) = k_{ij} = \mathbf{k}^T_{i,[j-1]} K_m^{-1}[j-1]\mathbf{k}_{[j-1],j}\,, \qquad (27)$$

where $t_i < \cdots < t_{j-m} < t_{j-m+1} < \cdots < t_{j-1} < t_j$. The condition (27) can be obtained from (23) by going to the matrix-vector notation . In (27) the following designations are used: $\mathbf{k}^T_{i,[j-1]} = \left(k_{i,j-m}, k_{i,[j-m+1]}, \ldots, k_{i,j-1}\right)$ - $m$ - dimensional row-vector values $k(s,t)$ at the points $T_m[j-1] = \{t_{j-m}, t_{j-m+1}, \ldots, t_{j-1}\}$ and in the point $t_i$ ;

$\mathbf{k}_{[j-1],j} = \left(k_{j-m,j}, k_{j-m+1,j}, \ldots, k_{j-1,j}\right)$ - $m$ - dimensional column-vector values $k(s,t)$ at the points $T_m[j-1]$ and in the point $t_i$ (in other words $\mathbf{k}_{[i],l}$ is $m$ -dimensional vector of covariance measurements at points $T_m[i]$ with measurement at point $t_l \in T_n$) ;

$K_m[j-1]$- ($m$ x $m$) covariance matrix of vector values of $Z(t)$ at the points $T_m[j-1]$, i.e. $K_m[j-1]$= $[k(t_s, t_l)](s, l = \overline{j-m, j-1})$.

A graphical illustration of notation to the formula (27) is shown on the Fig. 2.

Let in the points $T_n$ measured the $m$- connected Markov process $Z_M(t)$. Let $K_n$ be the covariance matrix of these measurements.

For the matrix $K_n$ vector $\mathbf{k}_{[i],j}(i \geq j)$ in (27) can be interpreted as a set of elements of $j$- th column of the $i$-$m$+1 to $i$ for $i \geq m$ or from 1 to $i$ under $i < m$. Thus, the dimension of the vector $\mathbf{k}_{[i],j}$ will be equal to $m$ under $i \geq m$ and will be equal $i$ for $i < m$. Matrix $K_m[i]$, thus, can be interpreted as a diagonal submatrix $K_n$, located at the intersection of rows and columns of the same name with numbers from $i$-$m$+1 to $i$ for $i \geq m$or from 1 to $i$ for $i < m$. Thus, the size of submatrix $K_m[i]$ will be equal $m$ x $m$ for $i \geq m$ and $i$ x $i$ for $i < m$.

Let the vectors $\mathbf{\Gamma}_i (i = \overline{1, n-1})$ defined as follows:

$$\mathbf{\Gamma}_i = K_m^{-1}[i]\mathbf{k}_{[i],i+1}\,, \; (i = \overline{1, n-1}). \qquad (28)$$

Obviously, the dimension of the vector $\mathbf{\Gamma}_i$ will be equal $m$ for $i \geq m$ and $i$ for $i < m$.

Taking into account the entered notation we can formulate the following

*Theorem 5*: 1) Covariance matrix $K_n$ for measurements at points $T_n$ of $m$- connected Markov process is a special case $A_n^m$ when $a_{ii} = k_{ii}(i = \overline{1, n})$, $\boldsymbol{\gamma}^T_{im} = \boldsymbol{\lambda}_{mi} = \boldsymbol{\gamma}_i(i = \overline{1, n-1})$, $A_{im} = A^T_{mi} = K_{im}$ where $K_{im}$- submatrices of the matrices $K_i(i = \overline{1, n})$, representing their right $m$ – column of a size $(i \times m)$ for $i \geq m$ and the size $(i \times i)$ for $1 < i < m$.

2) The matrix $K_n^{-m}$, inverse to $K_n^m$, is a band with a half-width band equal to $m$, which elements are defined by expressions

$$\alpha_i = k_{ii} - \mathbf{k}^T_{i,[i-1]} K_m^{-1}[i-1]\mathbf{k}_{[i-1],i} = k_{ii} - \mathbf{k}^T_{i,[i-1]}\mathbf{\Gamma}_{i-1}, (i = \overline{1, n}), \qquad (29)$$

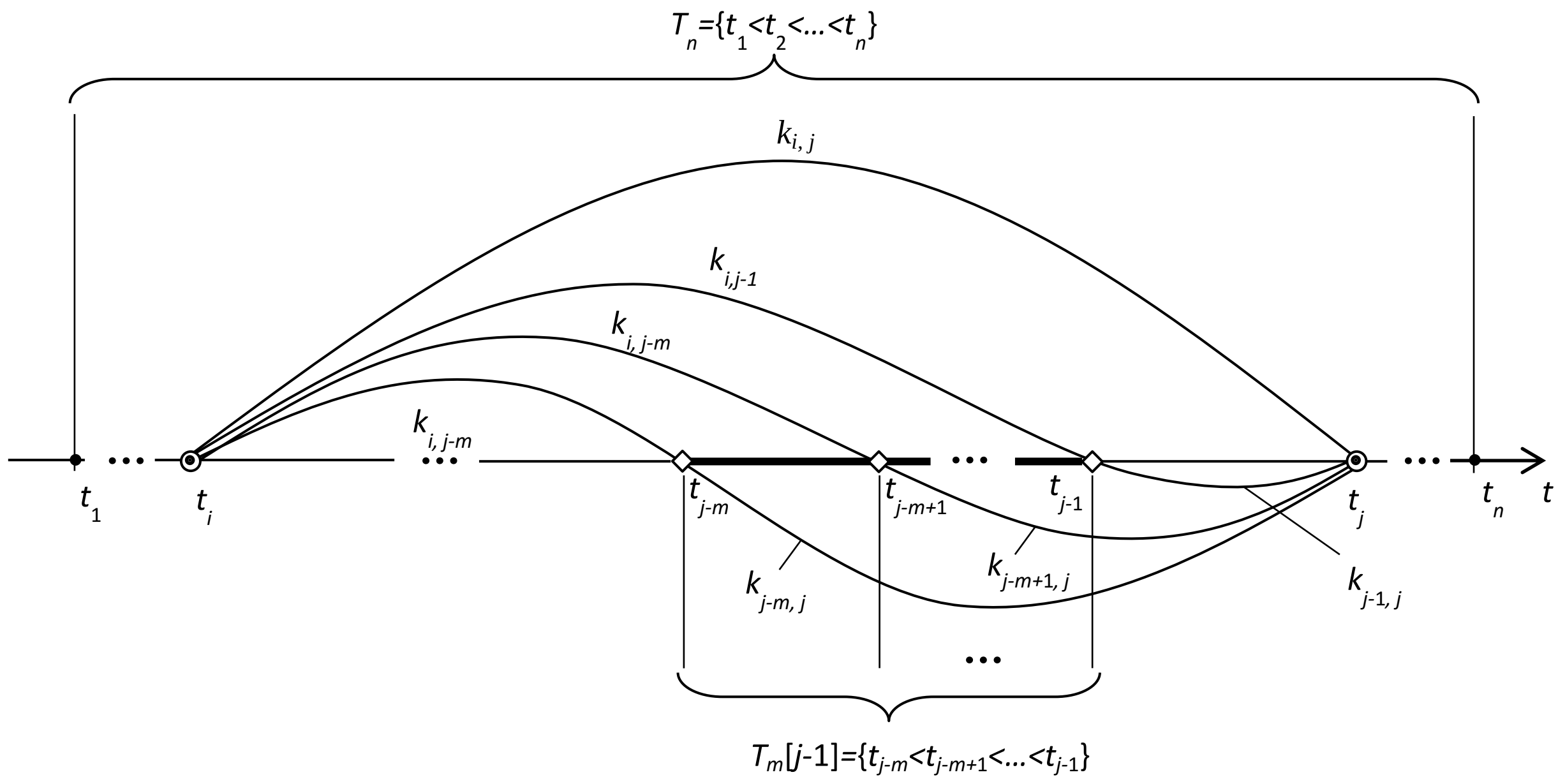

Fig. 2. Graphical illustration for the formula (27)

$$c_{ii} = \frac{1}{\alpha_i} + \sum_{k=0}^{w} \frac{\gamma^2_{m-k,i+k}}{\alpha_{i+k+1}}, (i = \overline{1,n}), \tag{30}$$

$$c_{i,i+k} = -\frac{\gamma_{i+k-1,m-k+1}}{\alpha_{i+k}} + \sum_{j=k}^{w} \frac{\gamma_{i+j,m-j}\gamma_{i+j,m+k-j}}{\alpha_{i+j+1}}, \; c_{i+k,i} = c_{i,i+k}, \tag{31}$$

$(i = \overline{1,n-1}; k = \overline{1,m})$,

where $w = m$-1 for $i \le n - m$ and $w = n - i$ - 1 for $i > n - m$.

*Note 11*: If in (30) and (31) the calculated upper limit value becomes smaller than the lower one, then the summarizing should not be executed, i.e. in $j,k > w$ and $j,k > m$ the second term on the right side of mentioned formula is identically equal to 0.

*C. Covariance function and the covariance matrix of the measurement of m- dimensional (vector) Markov process*

The matrix of covariance function $K(s,t)$ of the vector of Markov process $\mathbf{Z}_M(t)$ satisfies the conditions

$$K(s,t) = K(s,\tau)K^{-1}(\tau,\tau)K(\tau,t), \tag{32}$$

where $s < \tau < t$ or $s > \tau > t$. (Condition (32) is necessary and sufficient to determine a Markov process in a wide-sense.)

Let's now try to find a common view of the covariance matrix of measurement at the points $T_n$ of the vector Markov process.

Let $\mathbf{K}_N$ be a block covariance matrix measurement at points $T_n$ of $m$-dimensional vector process, consisting of $n^2$ blocks $K_{ij}(i,j = \overline{1,n})$. Blocks, in turn, represent the covariance matrix measurement of the size $m \times m$ of components of the vector $\mathbf{Z}_M(t)$.

Let's define the order of the square $m$ (sub) matrices $\Gamma_i \quad (i = \overline{1,n-1})$ and $\mathbf{\Gamma}_{ij} \quad (i,j = \overline{1,n-1})$ as follows:

$$\Gamma_i = K_{ii}^{-1}K_{i,i+1}, \tag{33}$$

$$\mathbf{\Gamma}_{ij} = \begin{cases} \Gamma_i\Gamma_{i+1}\ldots\Gamma_{j-1}\Gamma_j & for \quad j \ge i \\ \Gamma_j^T\Gamma_{j+1}^T\ldots\Gamma_{i-1}^T\Gamma_i^T & for \quad i \le j \end{cases} \; (i,j = \overline{1,n-1}). \tag{34}$$

Taking into account the entered designations it is possible formulate the following

*Theorem 6*: 1. Covariance matrix of $\mathbf{K}_N = \mathbf{K}_N^m$ measurements in points $T_n$ $m$-dimensional Markov process $\mathbf{Z}_M(t)$ is positive defined and is a special case of (18) for $A_{ii} = K_{ii}(i = \overline{1,n})$ , $\mathbf{\Lambda}_{ij} = \mathbf{\Gamma}_{ji}(i, j = \overline{1,n-1})$ . Elements $\mathbf{K}_N^m$ dependent (taking into account its symmetry) only on $(2n-1)m^2$ scalar quantities being the elements of sub-matrix $K_{ii}$ $(i = \overline{1,n})$ and $\Gamma_i(i = \overline{1,n-1})$ .

2. The inverse matrix $\mathbf{C}_N^m = \left(\mathbf{K}_N^m\right)^{-1} = \mathbf{K}_N^{-m}$ , divided into $(m \times m)$ blocks $C_{ij}^m$ , has a block-tridiagonal form with non-zero elements:

$$C_{ii}^m = A_{i+1}^{-1} M_i A_i^{-1}, \quad (i = \overline{1,n-1}), \quad C_{nn}^m = A_n^{-1}, \tag{35}$$

$$C_{i,i+1}^m = -\Gamma_i A_{i+1}^{-1}, \quad C_{i+1,i}^m = -A_{i+1}^{-1}\Gamma_i^T = C_{i,i+1}^m \quad , \tag{36}$$

where

$$A_i = K_{ii} - \Gamma_{i-1}^T K_{i-1,i-1}\Gamma_{i-1}, \quad (i = \overline{2,n}), \quad A_1 = K_{11}, \tag{37}$$

$$M_i = K_{i+1,i+1} - \Gamma_i^T \Gamma_{i-1}^T K_{i-1,i-1}\Gamma_{i-1}\Gamma_i, \quad (i = \overline{2,n-1}), \quad M_1 = K_{22} \tag{38}$$

The theorem 6 proofing is similar to the theorem 4 proofing.

## VI. An Example of Covariance Matrix Inversion for Vector Markov Process

The formula received above, despite of their external inconvenience, easy to use in practical calculations. We'll show it by the example of the inversion of the covariance matrix of 2D Markov process, when the covariance function of the analyzed process is given.

Below we give some examples of getting of common expressions for submatrices $K_{ii}$ $(i = \overline{1,n})$ , $K_{i,i+1}$ and $\Gamma_i$ $(i = \overline{1,n-1})$ of covariance matrix $\mathbf{K}_N^m$ of measurements at points $T_n$ for the 2D vector process. Also common expression for the sub-matrices was found $A_i\,(i = \overline{1,n})$ and $M_i\,(i = \overline{1,n-1})$ , with the help of which we can easily calculate the non-zero sub-matrices of inverse matrix $\mathbf{K}_N^{-m}$ , without resorting to the standard procedures of matrix inversion (formulas (33)-(38)).

Note that in the general case, the separate components of the vector Markov process may be a Markov and non-Markov. Relationships between the components can also be Markov, non-Markov and semi-Markov, i.e. when mutual covariance function $k_{ij}(s,t)$ component $Z_i$ and $Z_j$ $(i, j = \overline{1,m})$ *m*- dimensional vector process satisfy (23) for $s < \tau < t$ and not satisfy for $s > \tau > t$ .

*Example.* Let us consider a 2D non-stationary Markov process $\mathbf{Z}(t) = \left(Z_1(t), Z_2(t)\right)^T$ , defined by the covariance matrix function

$K(s,t) = \left\lfloor k_{ij}(s,t) \right\rfloor$ $(i, j = \overline{1,2})$ with elements

$$k_{11}(s,t) = \sigma_1^2 \min(s,t),$$

$$k_{22}(s,t) = \frac{\sigma_2^2}{2\alpha}\left[\exp(-\alpha\,|\,s-t\,|) - \exp(-\alpha(s+t))\right]$$

$$k_{12}(s,t) = \frac{\sigma_1\sigma_2}{\alpha}\begin{cases} \exp(-\alpha(t-s)) - \exp(-\alpha t) & for \quad s < t \\ 1 - \exp(-\alpha t) & for \quad s > t \end{cases},$$

$$k_{21}(s,t) = \frac{\sigma_1\sigma_2}{\alpha}\begin{cases} 1 - \exp(-\alpha s) & for \quad s < t \\ \exp(-\alpha(s-t)) - \exp(-\alpha s) & for \quad s > t \end{cases}$$

It is possible to verify that the matrix function $K(s,t)$ obey (32), and its elements obey (23), both at $s < \tau < t$ , and when $s > \tau > t$ . Thus, $\mathbf{Z}(t)$ and its components are Markov and Markov related processes in a wide-sense.

*Note 12* [10]. *For a normal distribution of centered component* $\mathbf{Z}(t)$ *coincides with the two-dimensional Markov process representing a solution for* $t_0 = 0$ *and* $Z_1(0) = Z_2(0) = 0$ *of system of stochastic differential equations (SDE )*

$$\frac{dZ_1(t)}{dt} = \sigma_1 N(t), \quad \frac{dZ_2(t)}{dt} = -\alpha Z_1(t) + \sigma_2 N(t),$$ *excitation by normal white noise* $N(t)$ *with unit variance.*

Let's write the expression for the submatrices $K_{ii}$ and $K_{i,i+1}$ $\left(t_{i+1} > t_i\right)$ (see.(33)) of the covariance matrix $\mathbf{K}_N$ ,

influencing on the formation of non-zero submatrices of inverse matrix $\mathbf{K}_N^{-m}$:

$$K_{ii}=\frac{1}{\alpha}\begin{bmatrix}\alpha\sigma_1^2 t_i & \sigma_1\sigma_2(1-\exp(-\alpha t_i)) \\ \sigma_1\sigma_2(1-\exp(-\alpha t_i)) & \sigma_2^2(1-\exp(-2\alpha t_i))/2\end{bmatrix},$$

$$K_{i,i+1}=\frac{1}{\alpha}\begin{bmatrix}\alpha\sigma_1^2 t_i & \sigma_1\sigma_2(\exp(-\alpha(t_{i+1}-t_i))-\exp(-\alpha t_{i+1})) \\ \sigma_1\sigma_2(1-\exp(-\alpha t_i)) & \sigma_2^2(\exp(-\alpha(t_{i+1}-t_i))-\exp(-\alpha(t_i+t_{i+1}))/2\end{bmatrix}$$

$(i=\overline{1,n-1})$.

In this sub-matrix $K_{ii}^{-1}$ will have the form:

$$K_{ii}^{-1}=\frac{1}{\alpha\cdot\det K_{ii}}\begin{bmatrix}\sigma_2^2(1-\exp(-2\alpha t_i))/2 & -\sigma_1\sigma_2(1-\exp(-\alpha t_i)) \\ -\sigma_1\sigma_2(1-\exp(-\alpha t_i)) & \alpha\sigma_1^2 t_i\end{bmatrix};\quad (i=\overline{1,n})$$ where

$$\det K_{ii}=\frac{\sigma_1^2\sigma_2^2}{\alpha}\left[\frac{t_i}{2}(1-\exp(-2\alpha t_i))-\frac{1}{\alpha}(1-\exp(-\alpha t_i))^2\right]$$

To present the matrix $\mathbf{K}_N$ in the form of $\mathbf{K}_N^m$ it is necessary to calculate sub-matrix $\Gamma_i$ (34):

$$\Gamma_i=K_{ii}^{-1}K_{i,i+1}=\begin{bmatrix}1 & 0 \\ 0 & \exp(-\alpha(t_{i+1}-t_i))\end{bmatrix}=\begin{bmatrix}\gamma_{1i} & 0 \\ 0 & \gamma_{2i}\end{bmatrix},\quad (i=\overline{1,n-1})$$ where $\gamma_{1i}=1,\quad \gamma_{2i}=\exp(-\alpha(t_{i+1}-t_i))$.

To calculate non-zero submatrices $\mathbf{K}_N^{-m}$ it is necessary previously calculate $A_i$ and $A_i^{-1}(i=\overline{1,n})$ (see formula (37)):

$$A_i=\frac{1}{\alpha}\begin{bmatrix}\alpha\sigma_1^2(t_i-t_{i-1}) & \sigma_1\sigma_2(1-\gamma_{2,i-1}) \\ \sigma_1\sigma_2(1-\gamma_{2,i-1}) & \sigma_2^2(1-\gamma_{2,i-1}^2)/2\end{bmatrix};\quad (i=\overline{2,n})$$

$$A_1=K_{11}=\frac{1}{\alpha}\begin{bmatrix}\alpha\sigma_1^2 t_1 & \sigma_1\sigma_2(1-\exp(-\alpha t_1)) \\ \sigma_1\sigma_2(1-\exp(-\alpha t_1)) & \sigma_2^2(1-\exp(-2\alpha t_1))/2\end{bmatrix};\quad (t_0=0)$$

It is easy to calculate

$$A_i^{-1}=\frac{1}{\alpha\det A_i}\begin{bmatrix}\sigma_2^2(1-\gamma_{2,i-1}^2)/2 & -\sigma_1\sigma_2(1-\gamma_{2,i-1}) \\ -\sigma_1\sigma_2(1-\gamma_{2,i-1}) & \alpha\sigma_1^2(t_i-t_{i-1})\end{bmatrix};\quad (i=\overline{2,n})$$ where

$$\det A_i=\frac{\sigma_1^2\sigma_2^2}{\alpha}\left[\frac{1}{2}(t_i-t_{i-1})(1-\gamma_{2,i-1}^2)-\frac{1}{\alpha}(1-\gamma_{2,i-1})^2\right];$$

$$A_1^{-1}=\frac{1}{\alpha\cdot\det A_1}\begin{bmatrix}\sigma_2^2(1-\exp(-2\alpha t_1))/2 & -\sigma_1\sigma_2(1-\exp(-\alpha t_1)) \\ -\sigma_1\sigma_2(1-\exp(-\alpha t_1)) & \alpha\sigma_1^2 t_1\end{bmatrix};\quad (i=\overline{1,n})$$

where

$$\det A_1=\frac{\sigma_1^2\sigma_2^2}{\alpha}\left[\frac{t_1}{2}(1-\exp(-2\alpha t_1))-\frac{1}{\alpha}(1-\exp(-\alpha t_1))^2\right]$$

Let's find a submatrix $M_i\,(i=\overline{1,n-1})$ (formula (38)), which together with the previously found submatrices $\Gamma_i$ and $A_i^{-1}$ determine all 3$n$-2 non-zero submatrices $C_{ii}\,(i=\overline{1,n})$ and $C_{i,i+1}=C_{i+1,i}\,(i=\overline{1,n-1})$ of the matrix $\mathbf{K}_N^{-m}$:

$$M_i=\frac{1}{\alpha}\left\|\begin{array}{c|c}\alpha\sigma_1^2(t_{i+1}-t_{i-1}) & \sigma_1\sigma_2(1-\gamma_{2,i-1}\gamma_{2i}) \\ \hline \sigma_1\sigma_2(1-\gamma_{2,i-1}\gamma_{2i}) & \sigma_2^2(1-\gamma_{2,i-1}^2\gamma_{2i}^2)/2\end{array}\right\|(i=\overline{2,n})$$

$$M_1=\frac{1}{\alpha}\left\|\begin{array}{c|c}\alpha\sigma_1^2 t_2 & \sigma_1\sigma_2(1-\exp(-\alpha t_2)) \\ \hline \sigma_1\sigma_2(1-\exp(-\alpha t_2)) & \sigma_2^2(1-\exp(-2\alpha t_2))/2\end{array}\right\|(t_0=0)$$

Simpler expressions are obtained for a uniform measurements $T_n^\tau$. Let $t_0$ = 0, $t_1$ = $\tau$, $t_2$ = 2$\tau$ and etc. Let $\sigma_1=\sigma_2=\sigma$. Then $\Gamma_1=\Gamma_2=\cdots=\Gamma_{n-1}=\Gamma;\ A_1=A_2=\cdots=A_n=A;\ M_1=M_2=\cdots=M_{n-1}=M$ and

$$A = \frac{\sigma^2}{\alpha}\begin{Vmatrix} \alpha\tau & (1-\gamma) \\ (1-\gamma) & (1-\gamma^2)/2 \end{Vmatrix}; \quad A^{-1} = \frac{\sigma^2}{\alpha \cdot \det A}\begin{Vmatrix} (1-\gamma^2)/2 & -(1-\gamma) \\ -(1-\gamma) & \alpha\tau \end{Vmatrix};$$

$$\Gamma = \begin{Vmatrix} 1 & 0 \\ 0 & \gamma \end{Vmatrix}; \qquad M = \frac{\sigma^2}{\alpha}\begin{Vmatrix} 2\alpha\tau & (1-\gamma^2) \\ (1-\gamma^2) & (1-\gamma^4)/2 \end{Vmatrix}$$

where $\gamma = \exp(-\alpha\tau)$:

$$\det A = \left(\sigma^4/\alpha^2\right)(1-\gamma)\left[\alpha\tau(1+\gamma)/2-(1-\gamma)\right].$$

Then $\mathbf{K}_N^{-m}$ can be represented as a block-tridiagonal matrix consisting of $n^2$ blocks of size $(m \times m)$:

$$\mathbf{K}_N^{-m} = A^{-1}\begin{Vmatrix} M & -A\Gamma & & & & \\ -\Gamma^T A & M & -A\Gamma & & \mathbf{0} & \\ & -\Gamma^T A & M & -A\Gamma & & \\ & & -\Gamma^T A & \ddots & & \\ & & & & -A\Gamma & \\ & \mathbf{0} & & -\Gamma^T A & M & -A\Gamma \\ & & & & -\Gamma^T A & A \end{Vmatrix} A^{-1}.$$

In contrast to the scalar case, for vector processes there are various possible ways of formation of the covariance matrix of the measurement (CMM). The above results relate to the case when CMM formed of $n^2$ submatrices of the size $(m \times m)$, representing measurements $m$ component of vector process in a given point. But it is possible to form a CMM so that it will consist of $m^2$ blocks of the size $(n \times n)$, representing measurement of one component of the field at the points $T_n^\tau$.

The general form of the matrix $K_N^{-m}$ for this case is shown below:

$$
K_N^{-m} = \frac{\sigma^2}{\alpha}
\begin{array}{c|cccccc|cccccc|}
 & 1 & 2 & 3 & \cdots & n-1 & n & 1 & 2 & 3 & \cdots & n-1 & n \\
\hline
1 & A_1 & A_2 & & & & & \Gamma_1 & \Gamma_2 & & & & \\
2 & A_2 & A_1 & A_2 & & \mathbf{0} & & \Gamma_3 & \Gamma_1 & \Gamma_2 & & \mathbf{0} & \\
3 & & A_2 & A_1 & \cdot & & & & \Gamma_3 & \Gamma_1 & \cdot & & \\
\vdots & & & \cdot & \ddots & \cdot & & & & \cdot & \ddots & \cdot & \\
n-1 & & \mathbf{0} & & \cdot & A_1 & A_2 & & \mathbf{0} & & \cdot & \Gamma_1 & \Gamma_2 \\
n & & & & & A_2 & A_1/2 & & & & & \Gamma_3 & 1 \\
\hline
1 & \Gamma_1 & \Gamma_3 & & & & & \Gamma_4 & \Gamma_5 & & & & \\
2 & \Gamma_2 & \Gamma_1 & \Gamma_3 & & \mathbf{0} & & \Gamma_5 & \Gamma_4 & \Gamma_5 & & \mathbf{0} & \\
3 & & \Gamma_2 & \Gamma_1 & \cdot & & & & \Gamma_5 & \Gamma_4 & \cdot & & \\
\vdots & & & \cdot & \ddots & \cdot & & & & \cdot & \ddots & \cdot & \\
n-1 & & \mathbf{0} & & \cdot & \Gamma_1 & \Gamma_3 & & \mathbf{0} & & \cdot & \Gamma_4 & \Gamma_5 \\
n & & & & & \Gamma_2 & 1 & & & & & \Gamma_5 & \Gamma_1/2
\end{array}
$$

where $A_1 = 2\alpha\tau$; $A_2 = -\alpha\tau$; $\Gamma_1 = 1-\gamma^2$; $\Gamma_2 = -\gamma(1-\gamma)$; $\Gamma_3 = -(1-\gamma)$; $\Gamma_4 = (1-\gamma^4)/2$; $\Gamma_5 = -\gamma(1-\gamma^2)/2$ i.e. matrix $K_N^{-m}$ consists of $m^2$ tridiagonal blocks of the size $(n \times n)$.

## VII. Conclusion

1. The convenient in usage forms of square matrix, whose inversions are tridiagonal, band or block-tridiagonal matrices, have been represented in this paper. In the work they are designated as $A_n^1$, where $n$ - dimension of a matrix, 1 – half-width of the band; $A_n^m$ where $m$ - half-width of the band, and $A_N^m$, where $N$ - the dimension of the matrix, $m$ - dimension of the blocks $(N = n \times m)$.

2. The objectives of the study of such matrices devoted a lot of work. A special feature of our work is that it shows that the recording of all three classes of matrices, you can use a common approach and a common (unique) matrix structure. Moreover, in the first case the matrix elements $A_n^1$ are formed of the scalar quantities, in the second case $A_n^m$ - of the dimensions of the vectors $m$, and in the third case $A_N^m$ of square blocks (submatrices) dimension $m$.

3. The simple inversion formulas of these matrices are founded, considering the peculiarities of the structure of the matrix. At the same time, the elements of inverse matrices depend only on:

- $3n-2$ elements included in the 3 central diagonal for the matrix $A_n^1$;
- $(2m+1)n - m(m+1)$ elements lying inside the band with the width 2m+1 (band's half-width equals *m*) for the matrix $A_n^m$;
- $(3n-2)(m \times m)$ elements for the matrix $A_N^m$ $(N = n \times m)$.

4. The savings were calculated in the amount of required memory and the number of operations in matrix inversion of the form $A_n^m$ relatively to a general matrix (fully completed).

5. It is shown that if the matrices $A_n^1$, $A_n^m$ and $A_N^m$ $(N = n \times m)$ are symmetric and positive definite, they are covariance matrices of measurements, respectively, simply (ordinary connected), multiply of the connectivity $m$ and $m$ - dimensional vector Markov processes in a wide-sense.

6. It is shown that the covariance matrix of the measurement (CMM) of *ordinary connected* Markov process in a wide-sense depends only on the variance value at the measuring points and the coefficients of the covariance between adjacent measurement points. Accordingly, for *multiply connected* Markov process CMM depends on the variance and coefficients of covariance between points standing from each other by an amount equal to or less than the connectivity of process $m$.

7. Calculations of revising covariance matrix for 2D Markov process and Markov connected components, showing applicability of received results for practical tasks solving.

8. The received results allow simplifying the solution of many problems of random processes statistics. In particular, dramatically simplifies the computational complexity of the estimating tasks, filtering and interpolation of random processes and fields using BLUE and GLSE, which have been using inverse of covariance matrix of measurements as a weighting matrix.

Due to the volume of the article, the article does not consider questions of approximation of an arbitrary random process Markov $m$-connected process; case sparse covariance matrices with no more $3n-2$ non-zero elements, arranged in arbitrary positions; getting recurrent algorithms and a number of other interesting problems of processing the results of measurements of random processes.

## APPENDIX

### PROOF OF THE THEOREM 1

The validity of Theorem 1 is shown by induction, using a recursive procedure by means of a serial (step-by step, successive) matrix bordering (Faddeev-Faddeeva bordering , see., [23]).

Firstly, let us consider the $i$-th step of the recurrent procedure of matrix inversion. According to [23], if

$$A_{i+1} = \left\| \begin{array}{c:c} A_i & \mathbf{a}_{i+1} \\ \hdashline \mathbf{a}_{i+1}^T & a_{i+1,i+1} \end{array} \right\|, \quad i = 1,2,3,\ldots$$

where $\mathbf{a}_{i+1}^T = (a_{i+1,1},\ldots,a_{i+1,i})$; $\mathbf{a}_{i+1} = (a_{1,i+1},\ldots,a_{i,i+1})^T$ - bordering a row vector and a column vector respectively of $i$-elements length,

$$A_{i+1}^{-1} = \left\| \begin{array}{c:c} A_i^{-1} + \dfrac{\mathbf{u}_{i+1}\mathbf{v}_{i+1}}{\alpha_{i+1}} \; {}_i & -\dfrac{\mathbf{u}_{i+1}}{\alpha_{i+1}} \\ \hdashline -\dfrac{\mathbf{v}_{i+1}}{\alpha_{i+1}} & \dfrac{1}{\alpha_{i+1}} \end{array} \right\|, \quad i = 1,2,3,\ldots \quad (38)$$

where $\alpha_{i+1} = a_{i+1,i+1} - \mathbf{a}_{i+1}^T\mathbf{u}_{i+1} = a_{i+1,i+1} - \mathbf{v}_{i+1}\mathbf{a}_{i+1}$; $\mathbf{u}_{i+1} = A_i^{-1}\mathbf{a}_{i+1}$; $\mathbf{v}_{i+1} = \mathbf{a}_{i+1}^T A_i^{-1}$.

Let the matrix $A_i^{-1}$ (size $i$ x $i$) found in the previous ($i$ - 1) - step recurrent procedure $(1 \le i \le n-1)$ has tridiagonal form corresponding to (6), and the $i$-dimensional vectors $\mathbf{a}_{i+1}$ and $\mathbf{a}_{i+1}^T$ :

$$\mathbf{a}_{i+1}^T = \left(\mathbf{a}_{i+1}^*\right)^T = \left[\prod_{l=1}^{i}\gamma_l a_{11}, \prod_{l=2}^{i}\gamma_l a_{22}, \ldots, \gamma_i a_{ii}\right];$$

$$\mathbf{a}_{i+1} = \mathbf{a}_{i+1}^* = \left[\prod_{l=1}^{i}\lambda_l a_{11}, \prod_{l=2}^{i}\lambda_l a_{22}, \ldots, \lambda_i a_{ii}\right]. \quad (39)$$

Carrying out the necessary calculations, we obtain the following formula:

$$\left.\begin{array}{l} \mathbf{u}_{i+1} = A_i^{-1}\mathbf{a}_{i+1} = A_i^{-1}\mathbf{a}_{i+1}^* = [0,0,\ldots,0,\lambda_i] = \mathbf{u}_{i+1}^* \\ \mathbf{v}_{i+1} = \mathbf{a}_{i+1}^T A_i^{-1} = \left(\mathbf{a}_{i+1}^*\right)^T A_i^{-1} = [0,0,\ldots,0,\gamma_i]^T = \mathbf{v}_{i+1}^* \end{array}\right\} \quad (40)$$

Substituting the values $\mathbf{u}_{i+1}$ and $\mathbf{v}_{i+1}$ ( ) to (38) and perform all necessary operations, we see that the matrix $A_{i+1}^{-1}$ will also be tridiagonal, the elements $A_{i+1}^{-1}$ correspond to (6). Thus, if the initial matrix, which begins the process of recurrent inverse is tridiagonal type and bordering vectors $\left\{\mathbf{a}_l, \mathbf{a}_l^T, l = \overline{i+1,n}\right\}$ are selected from the corresponding rows and columns (4), all subsequent matrices $\left\{A_i^{-1}, l = \overline{i+1,n}\right\}$, including the latest, are tridiagonal.

Successive (step-by-step) calculation $A_1^{-1}$, $A_2^{-1}$, $A_3^{-1}$ for the matrix $A_i^1$ $(i = \overline{1,3})$ shows that the matrix $A_3^{-1}$ is tridiagonal, i.e. the initial part of the procedure (A1) for the matrices (4) also leads to a tridiagonal matrix. This completes the proof of the theorem 1.

## References

[1] A. Asif,and J. M. F. Moura, "Block Matrices With L-Block-banded Inverse: Inversion Algorithms," *IEEE Trans. Signal Process.*, vol. 53, no. 2, Feb. 2005, pp. 630-642.

[2] A. Kavcic and J. M. F. Moura, "Matrix with banded inverses: algorithms and factorization of Gauss-Markov processes," *IEEE Trans. Inf. Theory*, vol. 46, no. 4, pp. 1495–1509, Jul. 2000.

[3] A. Asif and J. M. F. Moura, "Data assimilation in large time-varying multidimensional fields," *IEEE Trans. Image Process.*, vol. 8, no. 11, pp. 1593–1607, Nov. 1999.

[4] A. Asif,and J. M. F. Moura,, "Fast inversion of L-block-banded matrices and their inverses," in *Proc. IEEE Int. Conf. Acoust., Speech, Signal Process.*, vol. 2, Orlando, FL, May 2002, pp. 1369–1372.

[5] A. Sandryhaila and J. M. F. Moura " Eigendecomposition of block tridiagonal matrices," arXiv:1306.0217v1[math.SP]2 June 2013

[6] A. Asif and J.M.F. Moura, "Inversion of Block Matrices with Block Banded Inverses: Application to Kalman-Bucy Filtering," *Proc. IEEE Int. Conf. Acoust., Speech, Signal Process.*, vol. 1, Istanbul, Turkey, Jun 2000, pp. 608-611.

[7] A. Kavcic and J. M. F. Moura., "Information Loss in Markov Approximations," Tech. rep., Department of Electrical Engineering, Carnegie Mellon University, 1998. Manuscript of 30 pages, submitted to publication

[8] Дж. Л. Дуб, *Вероятностные процессы*. Москва: Иностранная литература, 1954. Translate from English: J.L. Doob. *Stochastic processes*. New York-John Wiley&Sons London-Chapman&Hall, 1953.

[9] Б.Р. Левин. *Теоретические основы статистической радиотехники*. 3-е издание. М: Радио и связь, 1989.

[10] В.С. Пугачев, И.Н. Синицын. *Стохастические дифференциальные системы (анализ и фильтрация)*. М.: Наука, 1990.

[11] Тихонов В.И., Миронов М.А. *Марковские процессы*. Москва: Советское Радио, 1977.

[12] Сосулин Ю.Г. *Теория обнаружения и оценивания стохастических сигналов*. Москва: Советское Радио, 1978.

[13] Rao C.R. "Estimation of parameters in linear model," *Ann.Statist.*, v.4, no. 6, pp. 1023-1037. 1976.

[14] Андерсон Т. *Статистический анализ временных рядов*. Москва: Мир, 1976. Translate from English: T.W. Anderson. The statistical analysis of time series. John Wiley&Sons. 1971.

[15] В.П. Ильин, Ю.И. Кузнецов. *Трехдиагональные матрицы.* Москва: Наука. 1985.

[16] Р. Хорн , Ч. Джонсон. *Матричный анализ.* Москва:Мир. 1989. R. Horn and C.R. Johnson. *Matrix analysis*. England. Cambridge University Press. 1986

[17] С. Писсанецки. *Технология разреженных матриц*. Москва: Мир, 1988. S. Pissanetzky, *Sparse Matrix Technology.* Academic Press Inc. 1984.

[18] У.Н. Бримкулов. Обобщенный метод наименьших квадратов в задачах планирования и анализа экспериментов при исследовании марковских случайных процессов . Дисс. . докт. техн. наук. М.:МЭИ, 1991. - 338 с. (in Russian).

[19] У.Н.Бримкулов. Обобщенный метод наименьших квадратов в задачах планирования и анализа экспериментов при исследовании марковских случайных процессов: автореферат диссертации на соискание ученой степени д-ра техн.наук:05.13.16. - М., 1991. - 40 с. Available: http://dlib.rsl.ru/loader/view/01000063973?get=pdf

[20] U. N. Brimkulov, "Peculiarities of the use of the generalized least squares method in problems of the identification of Markov random processes", *Avtomat. i Telemekh.*, 1991, no. 1, 69–78 (in Russian). English version: U.N. Brimkulov "Some features of generalized least squares method for identification of Markov stochastic processes," *Automation and Remote Control*. 1991. vol. 52, no.1, pp. 57-64.

[21] У.Н. Бримкулов. Ковариационная матрица измерений многосвязного марковского процесса и её применение в задачах оценивания случайных процессов. В сб.: Тезисы докладов IV Всесоюзной конференции «Перспективные методы планирования и анализа экспериментов при исследовании случайных полей и процессов». М, 1991. с. 64-65.

[22] У.Н. Бримкулов. Структура ковариационной матрицы измерений векторного марковского процесса для различных способов её формирования. В сб.: Тезисы докладов IV Всесоюзной конференции «Перспективные методы планирования и анализа экспериментов при исследовании случайных полей и процессов». М., 1991. с. 66-67.

[23] D.K. Faddev, V.N. Faddeva. *Computational methods of linear algebra*. Moscow: Fizmatgiz, 1963.